\documentclass[preprint,12pt]{elsarticle}
\usepackage{amsmath, mathrsfs, latexsym, amssymb, color, graphicx, comment, hyperref, scrhack}
\usepackage{lmodern}
\usepackage{physics}
\usepackage{amssymb}

\begin{document}

\begin{frontmatter}

\title{Composition heterogeneity influence on the phase formation dynamics in the Al--Y melt}

\author[label1,label2]{V.\,G.\,Lebedev}
\author[label1]{A.\,A.\,Lebedeva}
\author[label3]{M.\,G.\,Vasin}

\address[label1]{Department of Mathematics, Informatics and Physics, Udmurt State University, 426034 Izhevsk, Russia}
\address[label2]{Research center of metallurgical physics and materials science of UdmFRC UrB RAS, 426067 Izhevsk, Russia}
\address[label3]{Vereshchagin Institute of High Pressure Physics, Russian Academy of Sciences, 108840 Moscow, Russia}

\begin{abstract}
It is shown that at a description of a binary solution, in the presence of a liquid phase of variable composition and a stoichiometric solid phase, the concept of chemical potential can be introduced for stoichiometry, which qualitatively describes the dynamics of the re-distribution of the impurity at the contact of the phases. A model describing the slow dynamics of diffusion processes in an initially inhomogeneous sample is proposed. In the model under study, it is shown that the interaction of the impurity and the phase composition of the mixture, when deviating from equilibrium, leads to the development of instability, known as spinodal decay and mathematically described by the Cahn--Hilliard equation. Within the framework of this model system, a dispersion relation is constructed, from which the growth rate of unstable fluctuations from time is found and the influence of the model parameters on the instability value is investigated. The detected instability can explain the processes of slow non-monotonic relaxation that occurs when melting glass-forming metal alloys.
\end{abstract}


\end{frontmatter}


\section{Introduction}

Significant progress made in the study of the internal structure formation process in glass-forming materials does not help much in understanding the phenomenon of long-term, non-monotonic relaxation of in substances such as Al--Y or Al--Ni \cite{1, 2, 3, 4, 5}.  The main reason of this is that the mesoscopic scales at which major events develop are both too large for microscopic approaches, such as molecular dynamics, and too small (and too slow) for direct experimental investigation. For example, after melting aluminum with small additions of yttrium or nickel, the relaxation time can reach several hours. In metallurgy, this relaxation is explained by the processes of additional melting due to the slow dissolution of refractory solid inclusions in the melt. Changes in the internal structure of the melting substance are still quite a difficult task for the experiment, but there are problems in theory: the kinetics of such relaxation processes can not be explained in the framework of a linear diffusion model, the characteristic relaxation time in which should be on the order of a few seconds. The corresponding estimate of the characteristic dissolution time of the initial inhomogeneity Al$_3$Y with the characteristic size $10^{-5}$ m is made in \cite{VMI} and is equal to $10^{-2}$ s, which is significantly less than the relaxation times observed in the experiment: $\tau \approx~10^{4}$ s \cite{4,5}.

It was experimentally shown that slow non-monotonic viscosity decreasing was observed in Al--Y melts, both in the presence of other impurities and without them. It follows that the nature of this unusual phenomenon is mainly related to the relaxation features in the Al--Y melt. Therefore, for the sake of simplicity, we will limit ourselves to considering only the binary melt.

Unfortunately, in addition to this limitation and the fact of stable observation of this phenomenon, the experiment can not help anything yet. In the absence of theoretical and experimental clues, we can only hope for a plausible hypothesis that can shed light on the ongoing processes. This paper is devoted to the development of one of these hypotheses: we suppose the correlations of the composition of the impurity in the solution can affect the melting processes and the kinetics of the solution viscosity.

Seeing the marked slowness of the relaxation process, an analogy with the spinodal decay process of \cite{CH} is suggested as a possible explanation of the phenomenon in question, which was noted in \cite{VMI}. There were also some assessments in favor of the assumption made. Physically, the appearance of spinodal decay is associated with a characteristic type of dependence of the free energy (Gibbs potential) on the impurity concentration $x.$ When the Gibbs potential $G$ has one minimum concentration, the usual diffusion is observed in solutions. When an additional minimum appears (at some temperatures), the Gibbs potential region appears convex up. Since the impurity diffusion coefficient, up to a positive multiplier (mobility), is defined as the second derivative of the potential $D=M\partial^2 G/\partial x^2,$ it becomes negative in this region. A negative diffusion coefficient leads to instability, which results in a characteristic ``worm-like'' structure of the impurity distribution over space. The formation of the structure is associated with the formation of regions with different concentrations of impurity. However, according to thermodynamic data \cite{nims} (Fig.\,1) the Gibbs potential of the considered liquid solution is a smooth, convex-down function and cannot itself lead to any spinodal decay. Therefore, we can only talk about some effective process, the mathematical model of which is equivalent to the process of spinodal decay.
 \begin{figure}[htb]\label{fig2}
\begin{center}
\includegraphics[width=0.65\textwidth]{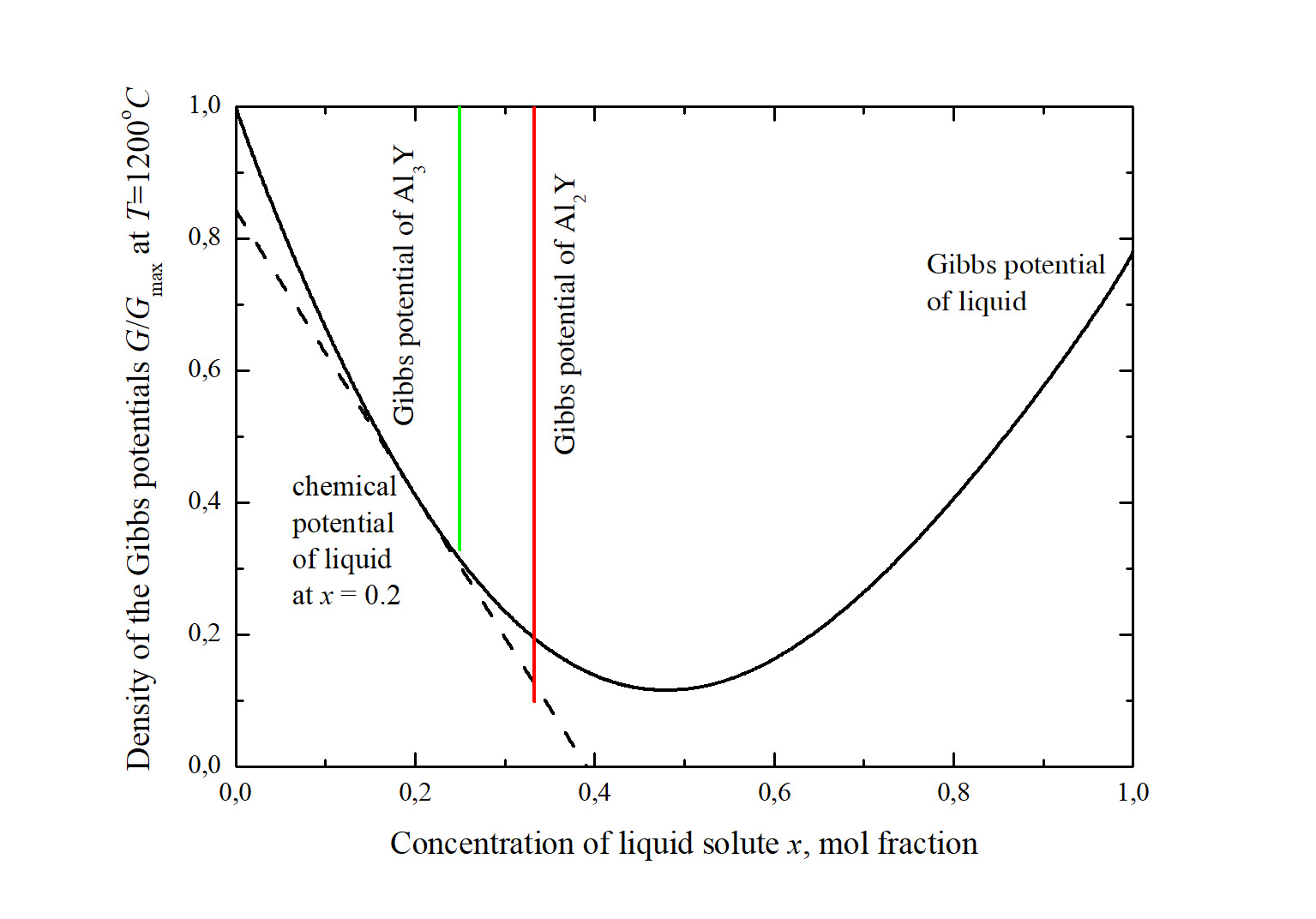}
\caption{
Real Gibbs potentials for liquid and two stoichiometric compounds Al$_3$Y and Al$_2$Y of the Al-Y solution at temperature $T=1473$K. The vertical lines are drawn for clarity, the energy of the stoichiometric phases corresponds to the lower points of the vertical lines}
\end{center}
\end{figure}

One can note another feature of this solution, shown in Fig.\,1: this is the presence of closely spaced vertical lines of stoichiometric compounds Al$_2$Y and Al$_3$Y registered in the initial samples. Therefore, it can be assumed that the presence of these compounds during melting may lead to local areas with an impurity concentration higher than the average for the sample. Moreover, it is well known that the physical processes occurring near peritectic are quite complex and interesting, and continue to be studied to the present day \cite{LFT,LUO}.

The purpose of this work is to study the processes of impurity redistribution between the liquid phase and stoichiometric compounds, which can result in instability characteristic of the spinodal decay process. However, since spinodal decay usually means separation into regions with different concentrations of impurity, but of the same aggregate nature, we will not talk further about spinodal decay, but about upward diffusion in the melting processes.
For simplicity, we will limit ourselves to a model, convex downwards potential, for which we can carry out analytical calculations to the end and analyze the dynamics of a system with liquid and stoichiometric phases.

\section{Model of liquid--solid phase transitions in the presence of a stoichiometric phase}

One of the challenges of describing a substance that has a stoichiometric phase (the phase of this constant composition) is the absence of the concept of the chemical potential of this phase. The phase of variable composition, for example for a binary solution, can be characterized by the molar density of the Gibbs potential $G(x)$ \cite{nims}. Then the chemical potential of the phase, or rather the difference in the chemical potentials of the components, can be obtained as a derivative \cite{Hill0}
$$\mu(x)=\frac{\partial G(x)}{\partial x}.$$
In the case of a stoichiometric phase, the Gibbs potential is defined by some function of temperature and does not depend on the composition of the phase, the molar concentration of the impurity in which is fixed and equal to $x_0.$ To get out of this situation, when calculating it is usually assumed that the Gibbs potential can be approximated by a very elongated parabolic function \cite{PGA,Hu}. This approach has its drawbacks and has been repeatedly criticized in the literature. We will approach this question from a slightly different angle, considering the stoichiometric phase as the limit of the phase of variable composition.

Let's start with a model isothermal problem of phase transformations in a binary melt with a molar impurity concentration $x$  in the liquid phase (phase $L$), assuming for simplicity that the melt at a given concentration corresponds to the volume density of the Gibbs potential $G^L(x)$ for alloys in which there is a non-monotonic relaxation of viscosity, the volume density of Gibbs potentials is a convex down the function of the concentration Fig.\,1. Therefore we limit ourselves to the simplest anharmonic approximation:
\begin{equation}
\label{Gl}
G^L=\frac12g_0(x-c_1)^2+\frac14f_0(x-c_2)^4,
\end{equation}
where $c_{1,2}$ is the position of the minimum $G^L (x)$ (Fig.\,2), and for simplicity $f_0=g_0.$
Next, we assume that this binary compound, in addition to the liquid phase, has a stoichiometric phase (phase $S$) of a fixed composition $x_0$, which takes a constant value in the volume of the phase and changes within the region where there is a mixture of phases.  For the stoichiometric phase, the Gibbs mole potential is a fixed value of $G^S_0$ at a given temperature (Fig.\,2).
\begin{figure}[htb]
\begin{center}
\includegraphics[width=0.45\textwidth]{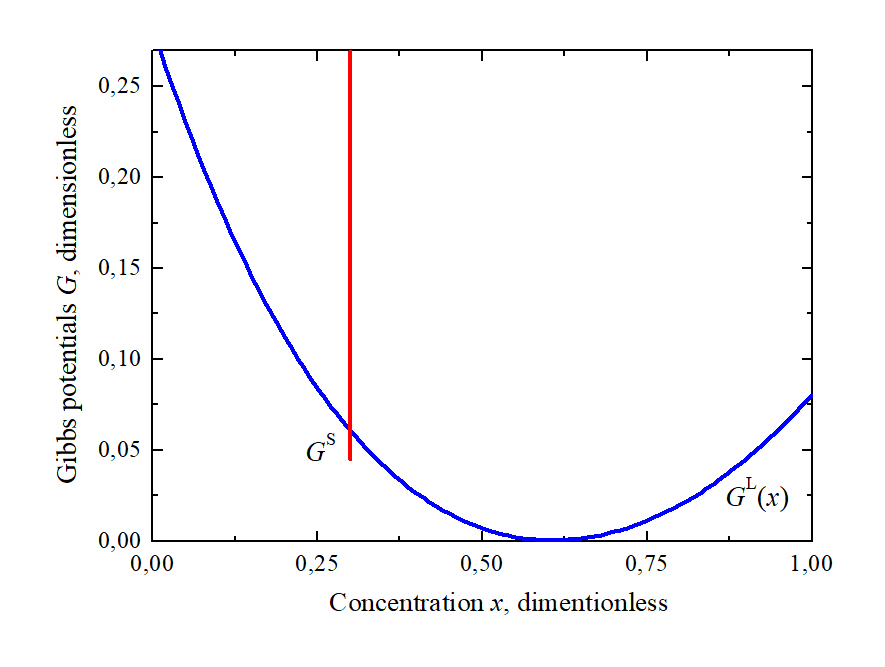}
\caption{Gibbs model potentials for liquid and stoichiometric coupling at $x_0=0.3$ }
\end{center}
\label{fig:3}
\end{figure}

To describe the phase state of the solution, we restrict ourselves to the scalar field $\varphi,$ such that for each unit of volume $\varphi$ corresponds to the fraction of the stoichiometric phase, and $(1-\varphi)$ is the fraction of the liquid phase. In the solid (stoichiometric) phase $S$ we assume that $\varphi=1,$ and in the liquid phase $L$ the phase field is zero.  In contrast to the well-known ideology of the phase field \cite{Elder}, we assume that the intermediate value of the field $0\le \varphi\le 1$ does not describe the interface between phases, but corresponds to a volume mixture of phases in the spirit of the quasi-equilibrium theory of crystallization \cite{Flem,LOV}. Ignoring the volume change during phase transformations and solidification, we write down the Gibbs potential in the form of phase interpolation. Since we are not interested in the interface, we omit the gradient contribution from the field $(\nabla\varphi)^2,$ in contrast to the phase field approach, but take into account the similar contribution for the concentration of the liquid phase $x,$ assuming the presence of correlations of the impurity concentration, but only in the liquid phase:
\begin{equation}
\label{GS}
\displaystyle G=\displaystyle\int\Big(\varphi G^S+(1-\varphi)G^L(x)
\displaystyle +\frac12\varepsilon^2\big(\nabla x(1-\varphi)\big)^2\Big)\,\mathrm{d}V,\\
\end{equation}
which formally exactly coincides with the Cahn--Hilliard free energy \cite{CH} at $ \varphi=0,$ but in this case, unlike \cite{CH}, the energy of the liquid phase is always convex down $\partial^2 G_L/\partial x^2\gg 0.$

Note that the concentration of the impurity in the liquid phase $x$ is not a constant value and can change both due to changes in the fraction of the phase and due to diffusion. Only the average concentration $\langle x\rangle$ is preserved, which can change in a unit of volume only due to diffusion flows $J_D$ in the liquid phase with a fraction $(1 - \varphi ):$
\begin{equation}
\label{cJ}
\frac{\partial}{\partial t}\langle x\rangle=-\nabla\cdot{\bf J}_D,
\end{equation}
which choice is due to the requirement to reduce the total Gibbs energy of the system in the relaxation processes.

The phase average concentration can be written as
\begin{equation}
\label{sx}
\langle x\rangle=\varphi x_0+(1-\varphi) x,
\end{equation}
differentiating which, formally considering the stoichiometric phase as a phase of variable composition, we find
\begin{equation}
\label{xt}
\varphi\dot{x}_0+(1-\varphi)\dot{x}=-\dot{\varphi}(x_0-x)-\nabla\cdot{\bf J}_D,
\end{equation}
where $\dot{\varphi}\equiv{\partial\varphi}/{\partial t}.$ This equation can be decomposed, following \cite{LG}, into equations for each of the phases separately. Formally, the sum of the equations
\begin{equation}
\label{xts}
\begin{array}{c}
\varphi\dot{x}_0=-\varphi\nabla\cdot{\bf J}_D-q,\\[12pt]
(1-\varphi)\dot{x}=-\dot{\varphi}(x_0-x)-(1-\varphi)\nabla\cdot{\bf J}_D+q,\\
\end{array}
\end{equation}
reproduces the original equation (\ref{xt}). The parameter $q$ will be defined further from the condition of constancy of the stoichiometric phase composition, and the contribution with $\dot{\varphi}$ also referred to the liquid phase for this reason.

Taking into account the equations (\ref{xt}), (\ref{xts}), we have a dynamic system of independent variables $x_0 ({\bf r}, t),$ $x({\bf r}, t),$ $\varphi({\bf r}, t),$ whose Lyapunov control functional is the total Gibbs energy of the system (\ref{GS}).

\section{Relaxation equation}

To obtain the relaxation equations of the dynamical system (\ref{GS}), we analyze the rate of change of the control potential, requiring a constant decrease in its value:
\begin{equation}
\begin{array}{c}
\displaystyle \frac{dG}{dt}=\displaystyle\int\Big(\dot{\varphi}\Big[G^S-G^L(x)+x\varepsilon^2\nabla^2\big(x(1-\varphi)\big)\Big]+\\[12pt]
\displaystyle+\dot{x}(1-\varphi)\Big[\mu_L-\varepsilon^2\nabla^2\big(x(1-\varphi)\big)\Big]+
\dot{x}_0\varphi\mu_S\Big)\mathrm{d}V\le 0,\\
\end{array}
\end{equation}
where both phases are considered as phases of variable composition
$$
\mu_S\equiv\frac{\partial G^S}{\partial x_0},\qquad\mu_L\equiv\frac{\partial G^L}{\partial x}.
$$
Taking into account the conservation law (\ref{xts}), the rate of change of the Lyapunov functional, after integration in parts, is rewritten as
\begin{equation}
\label{GSxt}
\begin{array}{c}
\displaystyle \frac{dG}{dt}=\displaystyle\int\Big(\dot{\varphi}\Big[G^S-G^L(x)-(x_0-x)\tilde{\mu}_L+\\[12pt]
\displaystyle+
x\varepsilon^2\nabla^2\big(x(1-\varphi)\big)\Big]+(\tilde{\mu}_L-{\mu}_S)q+\\[12pt]
\displaystyle+{\bf J}_D\nabla\Big[\varphi {\mu}_S+(1-\varphi)\tilde{\mu}_L\Big]\Big)\mathrm{d}V\le 0.\\
\end{array}
\end{equation}
where
\begin{equation}
\label{mus}
\tilde{\mu}_L=\mu_L-\varepsilon^2\nabla^2\big(x(1-\varphi)\big).
\end{equation}

The simplest choice that guarantees the decrease of the Lyapunov functional, in accordance with non-equilibrium thermodynamics \cite{RR}, is as follows:
\begin{equation}
\label{eq}
\begin{array}{c}
\displaystyle \dot{\varphi}=-M_{\varphi}\Big[G^S-G^L(x)-(x_0-x)\tilde{\mu}_L+
x\varepsilon^2\nabla^2\big(x(1-\varphi)\big)\Big],\\[12pt]
(\tilde{\mu}_L-{\mu}_S)q\le 0,\\[12pt]
\displaystyle {\bf J}_D=-M_D\nabla\Big[\varphi {\mu}_S+(1-\varphi)\tilde{\mu}_L\Big],\\
\end{array}
\end{equation}
where $M_D>0,$ $M_{\varphi}>0$ is the mobility factor.

Now that the driving forces and flows are determined, we take into account the constancy of the composition in the stoichiometric phase. To avoid changing $x_0$ in the equations (\ref{xts}), we use an arbitrary choice of the function $q.$ Choosing
$$
q= - \varphi\nabla\cdot{\bf J}_D,
$$
automatically we get $\dot{x}_0=0.$ Substituting $q$ in the equation for $\dot{x},$ with the definition (\ref{mus}) and the conservation law (\ref{xts}), the dynamics of the system with the control functional (\ref{GS}) will be determined by the follow equations:
\begin{equation}
\label{eq1}
\begin{array}{c}
\displaystyle \dot{\varphi}=M_{\varphi}\Big[G^L(x)-G^S+(x_0-x){\mu}_L
\displaystyle-x_0\varepsilon^2\nabla^2\big(x(1-\varphi)\big)\Big],\\[12pt]
\displaystyle (1-\varphi)\dot{x}=-(x_0-x)\dot{\varphi}
\displaystyle+\nabla\Big(M_D\nabla\Big[\varphi {\mu}_S+(1-\varphi)\tilde{\mu}_L\Big]\Big).
\end{array}
\end{equation}

\section{Equilibrium conditions and choice of chemical potential of the stoichiometric phase}

Under equilibrium conditions, the flows and rates of change of quantities turn to zero, so assuming that there is a uniform distribution of the concentration ($x^*=const$) and phases ($\varphi^*=const$), from (\ref{eq1}) we get
\begin{equation}
\label{eqr}
\begin{array}{c}
\displaystyle G^L(x^*)-G^S+(x_0-x^*){\mu}_L(x^*)=0,\\[12pt]
\displaystyle M_D\nabla\Big[\varphi^* {\mu}_S+(1-\varphi^*){\mu}_L(x^*)\Big]=0.\\
\end{array}
\end{equation}
Integrating the second equation, we find that over the entire space the following condition must be satisfied:
\begin{equation}
\label{eqr0}
\displaystyle \varphi^* {\mu}_S+(1-\varphi^*){\mu}_L(x^*)=const=\mu_0.
\end{equation}
For the condition (\ref{eqr0}) to be met for any $\varphi^*,$ just select
\begin{equation}
\label{eqr1}
\displaystyle {\mu}_S={\mu}_L(x^*)=\mu_0.
\end{equation}
Then from the first equation (\ref{eqr}) we find the equilibrium values $\mu_S=\mu_L(x^*)\!\!:$
\begin{equation}
\label{eqr2}
\displaystyle \mu_S=\mu_L(x^*)=\frac{G^S-G^L(x^*)}{(x_0-x^*)}.
\end{equation}

In it's meaning, the expression (\ref{eqr2}) defines the coefficient of the angular slope of the line connecting the Gibbs potential point of the stoichiometric phase and the Gibbs potential point with the current concentration $x$. This consideration suggests that the chemical potential of the stoichiometric phase in contact with the phase of variable composition $x,$ can be taken to be the expression (\ref{eqr2}) not only in equilibrium but also outside it:
\begin{equation}
\label{eqms}
\displaystyle \mu_S=\frac{G^S-G^L(x)}{(x_0-x)}.
\end{equation}
Thus, instead of equations (\ref{eqr1}), we finally have the following system of equations:
\begin{equation}
\label{eqo}
\begin{array}{c}
\displaystyle \dot{\varphi}=M_{\varphi}\Big[G^L(x)-G^S+(x_0-x){\mu}_L-
x_0\varepsilon^2\nabla^2\big(x(1-\varphi)\big)\Big],\\[12pt]
\displaystyle (1-\varphi)\dot{x}=-(x_0-x)\dot{\varphi}+\\[12pt]
\displaystyle+\nabla\bigg(M_D\nabla\bigg[\varphi \frac{G^S-G^L(x)}{x_0-x}+(1-\varphi)\bar{\mu}_L(x)\bigg]\bigg).\\
\end{array}
\end{equation}

Let's qualitatively analyze the change in the composition of the solution in contact with the phase of variable composition in this model. Obviously, with this choice of the chemical potential (\ref{eqr2}) of stoichiometry, taking into account the Gibbs potential of the phase of variable composition (\ref{Gl}), there are three fixed points, relative to the map given by the diffusion equation in the relations (\ref{eq1}):
\begin{itemize}
	\item $x=x_0,$ unstable fixed point;
	\item $x\!=\!x^*_{1,2},$ stable fixed point, the concentration of which is determined by the following equation: $$G^L(x^*)-G^S+(x_0-x^*){\mu}_L(x^*)=0.$$
\end{itemize}

Some graphs of Gibbs model potentials for liquid and stoichiometric phases are shown in Fig.\,3. in this figure, in addition to the convex downwards Gibbs potential of the liquid phase and the Gibbs potential of stoichiometry, shown for convenience by a vertical line, the lines of chemical potentials of the phases are drawn. From the ratio of chemical potentials, it can be seen that at a given initial composition of the liquid phase $x<x_0,$ it begins to receive (or give away an impurity) from the stoichiometric phase, approaching the point $A$ over time. We draw a line corresponding to $\mu_S$ at some point $x=x_1.$ Due to the convexity of the potential $G^L$, this line will also pass through the point $x_2.$ Comparing the slope of this line with the tangents at points $x=x_{1,2},$ we come to the statement about the existence of the point $A.$ When $x>x_0$ everything is mirrored.
\begin{figure}[htb]
\begin{center}
\includegraphics[width=0.6\textwidth]{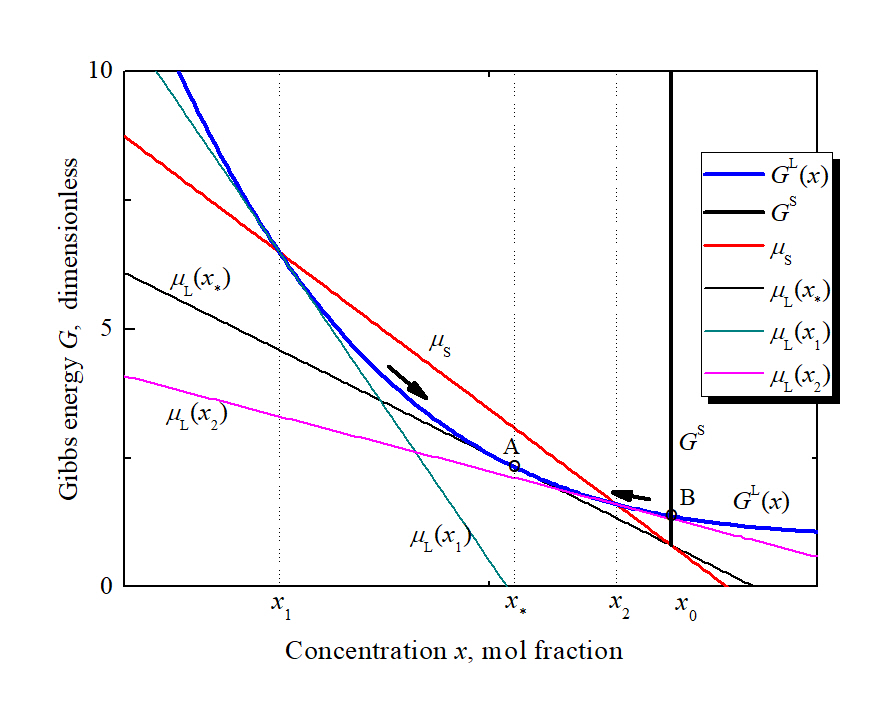}
\caption{Gibbs potentials of the liquid and stoichiometric phases, and the existence of equilibrium points.}
\end{center}
\label{fig:4}
\end{figure}

For more clarity, let's consider the behavior of chemical potentials of phases, shown in Fig.\,4. The singularity in the chemical potential of stoichiometry corresponds to the gap of the function (point $B,$) which necessary to ensure the existence of points not only of phase but also of diffusion equilibrium (points $A,$ $C$). Indeed, from Fig.\,4 it follows that in the region to the left of the point $A$ the chemical potential of the stoichiometry is higher than the chemical potential of the liquid phase. Therefore, the impurity should go from stoichiometry to the liquid phase, which enrichment leads to the fact that the condition of the liquid phase will be described by a point approaching the point $A$ to the left of the line of the liquid phase chemical potential. In the $A$--$B$ range the chemical potential of the liquid phase prevails, which causes the impurity to leave the liquid passing to the stoichiometric phase. In this case, the liquid is impoverished by the impurity and its state approaches the $A$ point on the right.
\begin{figure}[htb]\label{fig:5}
\begin{center}
\includegraphics[width=0.6\textwidth]{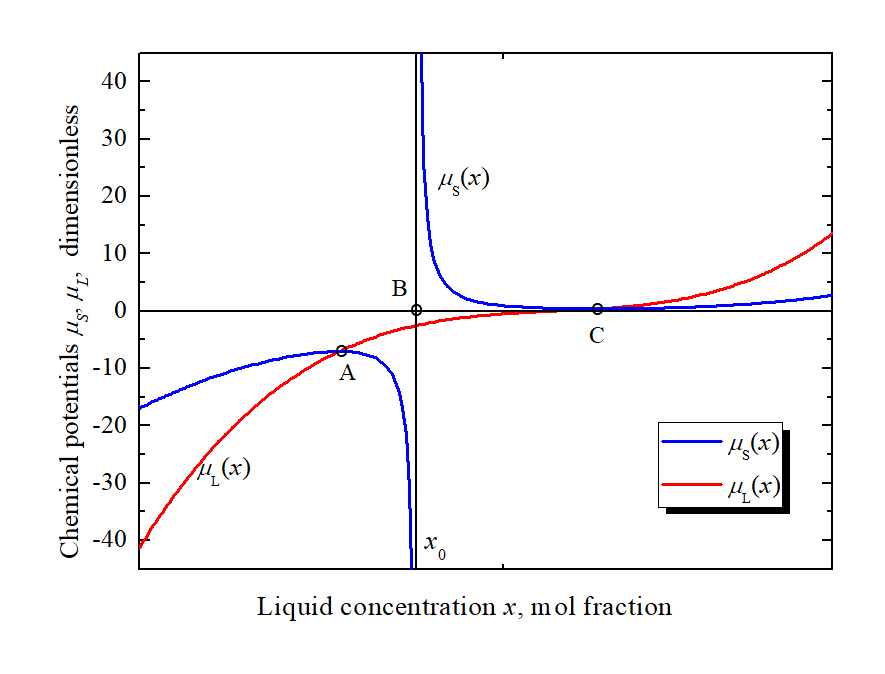}
\caption{Dimensionless chemical potentials of phases depending on the impurity concentration in the liquid phase. Point $B$ is the discontinuity of the chemical potential of the stoichiometric phase at $x=x_0.$ Points $A,$ and $C$ is the equilibrium points (for equality of chemical potentials $\mu_S=\mu_L$)}
\end{center}
\end{figure}

The analysis shows that despite the arbitrary choice of the chemical potential of the stoichiometric phase, its introduction (\ref{eqms}) leads to a fairly plausible behavior of diffusion processes in the presence of a steochiometric phase.

\section{Linearized equations}

To simplify further analysis of dynamics, we will move in the equations (\ref{eq1}) to dimensionless variables.
Counting them as constants and allocating characteristic scales over space, as some distance $L,$ and over time, as
$$t_0=\big(g_0M_{\varphi}\big)^{-1},$$
after replacing
$$\tilde{G}^S=g^{-1}_0{G^S},\quad \tilde{G}^L=g^{-1}_0{G^L},$$
$$\tilde{\mu}_S=g^{-1}_0{\mu_S},\quad \tilde{\mu}_L=g^{-1}_0{\mu_L},$$
we come to equations of the following form:
\begin{equation}
\label{eq2}
\begin{array}{c}
\displaystyle \dot{\varphi}=\tilde{G}^L-\tilde{G}^S+(x_0-x)\tilde{\mu}_L-
x_0\sigma\nabla^2\big(x(1-\varphi)\big),\\[12pt]
\displaystyle (1-\varphi)\dot{x}=-(x_0-x)\dot{\varphi}+
D_0\nabla^2\Big[\frac{\varphi}{x_0-x}\Big(\tilde{G}^S-\tilde{G}^L\Big)+\\[12pt]
\displaystyle+(1-\varphi)\big(\tilde{\mu}_L-\sigma\nabla^2\big(x(1-\varphi)\big)\big)\Big],\\
\end{array}
\end{equation}
where $\sigma=\varepsilon^2/g_0L^2,$ $D_0=M_D/L^2M_{\varphi}.$ Thus, only two dimensionless complexes (parameters) that characterize correlations ($\sigma$) and the contribution of diffusion ($D_0$), remain in equations. As mentioned earlier, for simplicity, the potential parameter $f_0$ is set to $g_0.$

As the initial state, we choose a state close to the equilibrium state, for which $\varphi^*=const$ and $x^*=const$ satisfy the following condition:
$$\mu_S(x^*)=\mu_L(x^*).$$

For small deviations from the initial state $\varphi ({\bf r}, t)=\varphi^*+ \delta\varphi({\bf r}, t)$
and $x({\bf r}, t)=x^*+\delta x({\bf r}, t),$ given that
$$(1 - \varphi)x\approx (1-\varphi^*)\delta x-x^*\delta\varphi,$$
we get the linearization in the form
\begin{equation}
\label{eq3}
\begin{array}{c}
\displaystyle \dot{\delta\varphi}=\Big(x^*\nabla^2\delta\varphi-(1-\varphi^*)\nabla^2\delta x\Big)x_0\sigma+
(x_0-x^*)\frac{\partial\tilde{\mu}_L(x^*)}{\partial x^*}\delta x,\\[12pt]
\displaystyle (1-\varphi^*)\dot{\delta x}=-(x_0-x^*)\dot{\delta\varphi}+D_p\nabla^2\delta x-\\[12pt]
\displaystyle -(1-\varphi^*)D_0\sigma\nabla^4\Big[(1-\varphi^*)\delta x-x^*\delta\varphi\Big],\\
\end{array}
\end{equation}
where
$$D_p=\bigg[\varphi^*\frac{\partial\tilde{\mu}_S(x^*)}{\partial x^*}+(1-\varphi^*)\frac{\partial\tilde{\mu}_L(x^*)}{\partial x^*}\bigg]D_0.$$

Let us represent $\tilde{G}^S$ in terms of the deviation from the Gibbs potential of the liquid phase at $x_0\!:$
$$\tilde{G}^S=\tilde{G}^L(x_0)-\Delta G,$$
then
$$\tilde{\mu}_S\equiv \tilde{\mu}_S(\Delta G)\quad D_p=D_p(\Delta G).$$
Excluding ${\delta\varphi}$ from the diffusion equation, we get:
\begin{equation}
\label{eq4}
\begin{array}{c}
\displaystyle \dot{\delta x}=\tilde{D}_p\nabla^2\delta x-(1-\varphi^*)D_0\sigma\nabla^4\delta x-\\[12pt]
\displaystyle -\frac{(x_0-x^*)^2}{(1-\varphi^*)}\frac{\partial\tilde{\mu}_L(x^*)}{\partial x^*}\delta x-\frac{x_0-x^*}{1-\varphi^*}x^*x_0\sigma\nabla^2\delta\varphi+\\[12pt]
\displaystyle +D_0\sigma x^*\nabla^4\delta\varphi,\\
\end{array}
\end{equation}
where
$$\tilde{D}_p=(x_0-x^*)x_0\sigma+\frac{D_p}{(1-\varphi^*)}.$$

The resulting equation formally corresponds to the Cahn--Hilliard equation with a source that depends on $\varphi.$ Since the Cahn--Hilliard equation describes the development of instability in the distribution of impurity in the volume of solution due to the negative diffusion coefficient for the second derivative in the equation, we investigate the dynamics of small deviations from equilibrium in our model system.

\section{Fourier analysis of deviations from equilibrium}

Now we consider the dynamics of small deviations from the equilibrium position over time $t,$ assuming the presence of small fluctuations of the phases ${\delta\varphi}_0$ and the composition ${\delta\xi}_0$ at the initial time. For simplicity, we restrict ourselves to a one-dimensional infinite domain whose points of space are numbered by the coordinate $z,$ assuming that
\begin{equation}
\label{fo1}
\left\{
\begin{array}{c}
\displaystyle {\delta\varphi}=C_1e^{\Omega t+iqz},\\[12pt]
\displaystyle {\delta\xi}=C_2e^{\Omega t+iqz},\\
\end{array}
\right.
\end{equation}
where $\Omega$ is the frequency, $q$ is the wave number.

Substituting (\ref{fo1}) into the equations (\ref{eq3}), we find the dispersion relation
$$
\begin{array}{c}
\displaystyle \left(\Omega+\tilde{D}_p q^2+(1-\varphi^*)D_0\sigma q^4+\frac{(x^*-x_0)^2}{1-\varphi^*}\frac{\partial\tilde{\mu}_L(x^*)}{\partial x^*}\right)\times\\[12pt]
\displaystyle \times\Big(\Omega+x^*x_0\sigma q^2\Big)-\sigma x^*q^2\left(D_0q^2+\frac{(x_0-x^*)x_0}{1-\varphi^*}\right)\times\\[12pt]
\displaystyle \times\Big(x_0-x^*+\sigma x_0q^2(1-\varphi^*)\Big)=0,\\
\end{array}
$$
which roots, $\Omega_p$ and $\Omega_m$, are shown further in Fig.\,5 and Fig.\,6 for different values of the parameters $\Delta G,$ for a fixed value $\sigma=0.01.$

For model calculations, the following parameters were accepted: $D_0=1,$ $x_0=0.3,$ $\varphi^*=0.25.$ in our model problem, the temperature is not explicitly present. As a parameter that determines the mutual location of Gibbs potentials, we use the parameter $\Delta G,$ which plays the role of some "'effective temperature". Another parameter that determines the dynamics of small deviations from equilibrium is the correlation between the diffusion $D_0$ and the contribution of correlations $\sigma.$ Since $D_0$ is chosen as a unit ($D_0=1$), then we analyze the dynamics of small deviations from the equilibrium at $\sigma=0\div 1.$
\begin{figure}[htb]
\begin{center}
\includegraphics[width=0.45\textwidth]{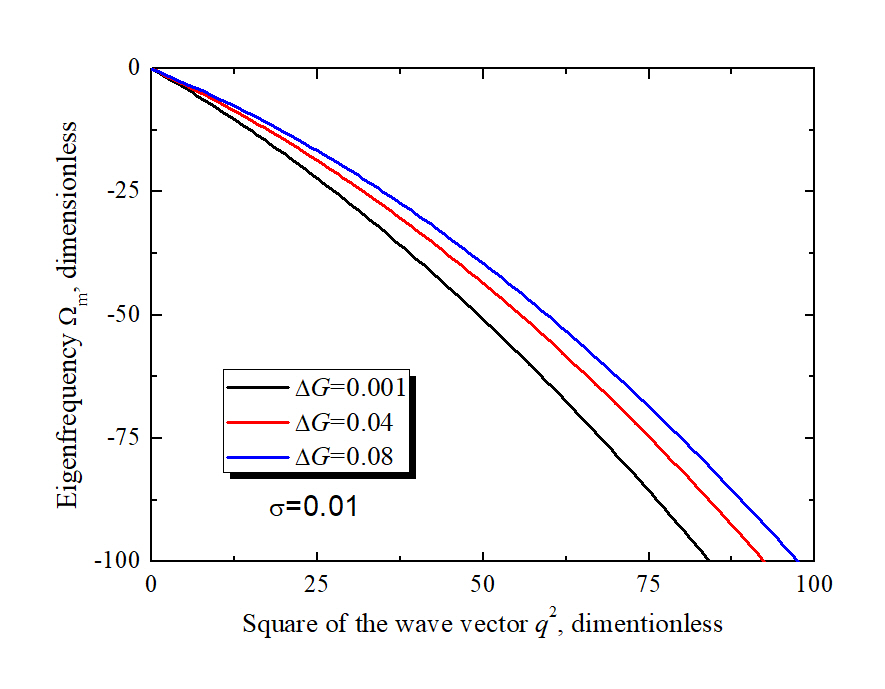}
\caption{Dispersion dependence of $\Omega_m$ on the square of the wavenumber for $\sigma=0.01$ and various values of $\Delta G$}
\end{center}
\label{fig:6}
\end{figure}

\begin{figure}[htb]
\begin{center}
\includegraphics[width=0.45\textwidth]{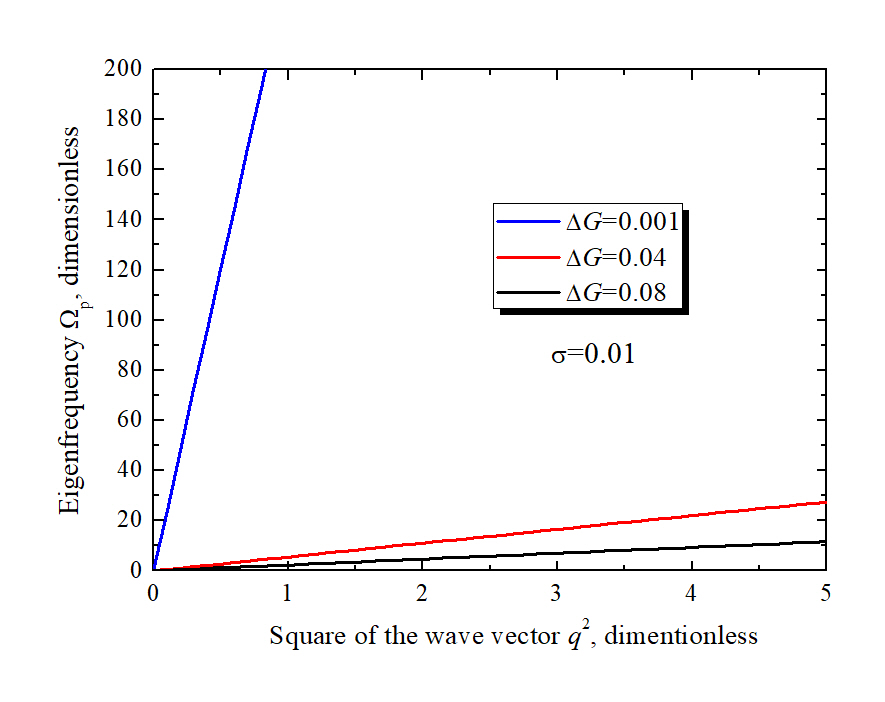}
\caption{Dispersion dependence of $\Omega_p$ on the square of the wavenumber for $\sigma=0.01$ and various values of $\Delta G$}
\end{center}
\label{fig:7}
\end{figure}

The most important conclusion that can be drawn here from Fig.\,5 and Fig.\,6 is that $\Omega_p(q^2)$ for possible $\Delta G$ causes instability of the corresponding mode. Graph of $\Omega_p(q^2)$ increases for small $q$ on the considered scales. This means that the instability when deviating from the equilibrium point is manifested at least on a sufficiently large scale of lengths. The fast development of perturbations access on small scales and slow on large ones. Given that the eigenvector corresponding to $\Omega_p(q^2)$ is a superposition of the phase field and concentration field, it is clear that each of these variables will have both and unstable and stable parts. Therefore, at small time intervals $t \approx 1/\Omega_p(q^2),$ the inhomogeneity of the distribution will appear for both the concentration and the phase field.

Another conclusion can be drawn from the graph in Fig.\,6, if we return to the interpretation of $\Delta G$ as an ``effective temperature''. At the temperature at which the Gibbs stoichiometry potential is close to (just below) the liquid phase potential, the most unstable state is observed ($\Omega_p (q^2)>0$).  With a further decrease in the ``effective temperature'' (an increase in $\Delta G$), the instability is significantly reduced. For a real system, this behavior can be interpreted as strong fluctuations in inhomogeneity of concentration and the nuclei of a new phase in the liquid with weak supercooling. As the system continues to cool down, the growth rate decreases. But in this case, deviations from the equilibrium are no longer small. Therefore such a study goes beyond the framework of the presented model and requires of dynamic analysis of the resulting perturbations not only at small times.

Since the dependency $\Omega_m(q^2)$ qualitatively differ little from each other, causing a rapid attenuation of the corresponding eigenvalue of deviations from equilibrium. Further, we will limit ourselves to analyzing the dependence on the parameters only of the eigenvalue $\Omega_p(q^2)$ (Fig.\,7--8).

\begin{figure}[htb]
\begin{center}
\includegraphics[width=0.45\textwidth]{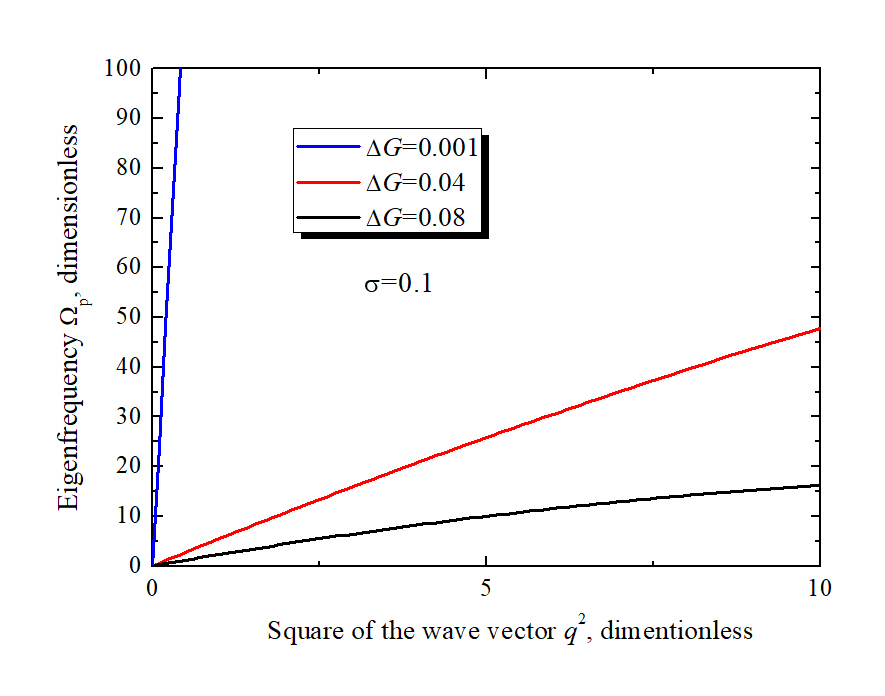}
\caption{Dispersion dependence of $\Omega_p$ on the square of the wavenumber for $\sigma=0.1$ and various values of $\Delta G$}
\end{center}
\label{fig:8}
\end{figure}

\begin{figure}[htb]
\begin{center}
\includegraphics[width=0.45\textwidth]{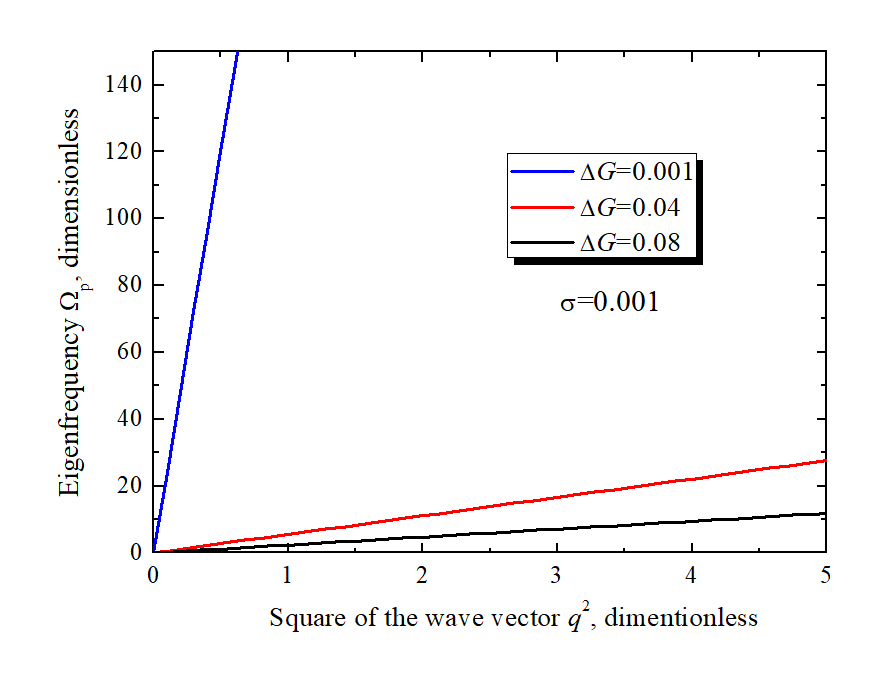}
\caption{Dispersion dependence of $\Omega_p$ on the square of the wavenumber for $\sigma=0.001$ and various values of $\Delta G$}
\end{center}
\label{fig:9}
\end{figure}

One can see that the dependencies $\Omega_p(q^2)$ in these drawings are qualitatively similar, and differ only in scaling along the $q^2$ axis. It is enough to compare the parameter values on this axis. Also note that in the absence of a correlation contribution ($\sigma=0$), the
instability does not occur, for any values $q^2$ the value $\Omega_p (q^2)<0.$ Thus, it should be concluded that not only spinodal decay forces to take into account the correlations of the composition in the liquid, but also the formation of stoichiometry. The latter is purely formal, from the point of view of mathematics, allows us to draw an analogy between spinodal decay and eutectic formation. The previous figures show the dependencies for small values of the wave vector. To illustrate the behavior of the unstable mode, we present a graph in Fig.\,9, which demonstrates the appearance of the instability maximum. On other charts, this behavior is not visible, because the scale is stretched along the axis of the wave vector $q^2.$

\begin{figure}[htb]
\begin{center}
\includegraphics[width=0.45\textwidth]{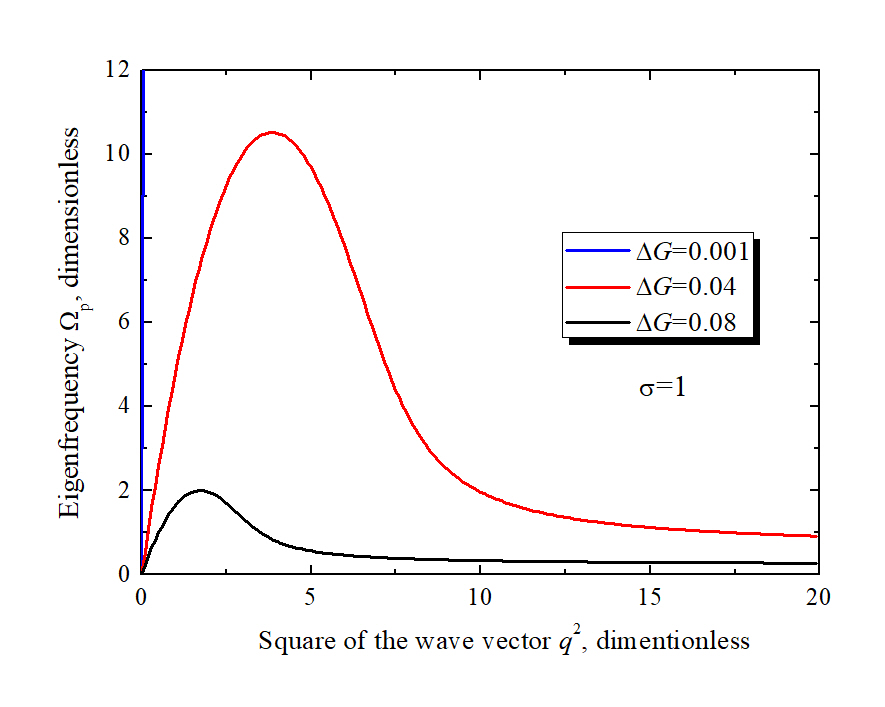}
\caption{Dispersion dependence of $\Omega_p$ on the square of the wavenumber for $\sigma=1$ and various values of $\Delta G$}
\end{center}
\label{fig:10}
\end{figure}

\section{Conclusions}

Thus, the model describing the processes of melting and solidification in a two-phase system consisting of a liquid and stoichiometric phase is proposed. The model is based on the introduction of an order parameter for the phase state and the chemical potential of the stoichiometric phase. The resulting equations have diffusive instability under certain conditions and behave similarly to the Cahn--Hilliard equation. The Fourier analysis of small deviations from the equilibrium position allowed out the expressions for the structural factor and the instability growth index. Their analysis of which shows that in the presence of a stoichiometric phase in solutions, ascending diffusion processes similar to spinodal decay can occur.

Note that the mechanism of upward diffusion considered in this paper is physically different from the Cahn--Hilliard spinodal decay \cite{CH}. If in spinodal decay the instability is associated with the upward bulge of the potential, then in our case the instability is due to the redistribution of the impurity between the phases. If at some point in time an impurity concentration locally occurs in the solution that exceeds some critical value, an excess of this concentration leads to the formation of a stoichiometric phase that takes a well-defined part of the impurity. The excess impurity is displaced into the liquid phase, leading to an increase in instability. Formally, mathematically, the model of such a process turns out to be quite equivalent to the Cahn--Hilliard model, although, as noted earlier, the potential of the liquid phase always remains a convex down function.

Another consequence of this study is a qualitative confirmation of the possible origin of long-term relaxation and non-monotonic behavior of viscosity in melting processes, based on the use of the Cahn--Hilliard equation \cite{VMI}. Of course, this work is based on a model system, so the resulting picture needs both additional theoretical research on real materials and additional experimental data, preferably obtained on matrix x-ray structures. This equipment can provide data on the growth rate of the structural factor during the melting of solid solutions of Al--Y and Al--Ni.

\section*{Acknowledgments}

The work was supported by Russian Foundation for Basic Research, Grants 18-02-00643 (MV) and 18-42-180002 (VL). Part of the work was carried out within the framework of the state assignment of the Ministry of Education and Science of Russia (No.AAAA-A17-117022250039-4)


\end{document}